\documentclass[fleqn,twoside]{article}
\usepackage{espcrc2}
\RequirePackage{lineno}

\usepackage{graphicx}
\usepackage[figuresright]{rotating}

\newcommand{\AmS}{{\protect\the\textfont2
  A\kern-.1667em\lower.5ex\hbox{M}\kern-.125emS}}

\title{Observation of the anisotropy in arrival direction of Cosmic Rays with IceCube}                                                                                                   

\author{S. Toscano \address[email]{E-mail: toscano@icecube.wisc.edu} for the IceCube Collaboration \address[UWMadison]{http://icecube.wisc.edu}\\                                                                              
IceCube Research Center, University of Wisconsin, \\                                                                                                                               
Madison, WI 53703, U.S.A.  }                                                                                                                                 

\begin{document}                                                                                                                                                                   
\begin{abstract}                                                                                                                                                                   
The IceCube Neutrino Observatory is a kilometer-scale detector currently under construction at the South Pole.                                                                     
In its final configuration the detector will comprise 5160 Digital Optical Modules (DOMs) deployed on 86 strings between 1.5-2.5 km deep within the ice.                                  
While still incomplete, the detector has already recorded tens of billions of cosmic ray muons with a median energy of 20 TeV.                                                                                                                                               
This large sample has been used to study the arrival direction distribution of the cosmic rays.                                                                                    
We report the observation of an anisotropy in the cosmic rays arrival direction at two different angular scales.                                                                           
The observed large scale anisotropy seems to be a continuation of similar structures observed in the Northern Sky by several experiments. IceCube observes also significant features on the angular scale of ~20$^\circ$ - 30$^\circ$ that might be part of the larger scale structure.

\vspace{1pc}
\end{abstract}

\maketitle

\section{INTRODUCTION}

During the last twenty years several underground experiments have demonstrated the presence, in the northern sky, of an anisotropy in the arrival direction of the cosmic rays (CRs) at energies up to few hundred GeV \cite{Nagashima89}, \cite{Nagashima98}. Recently, both underground and surface array experiments have observed a sidereal anisotropic modulation on the order of 10$^{-4}$ in the multi-TeV range ( \cite{Amenomori06}, \cite{ARGO09}, \cite{Guillian07} and \cite{Abdo09},). The origin of the anisotropy is unknown and observations of galactic cosmic ray anisotropy at different energy and angular scales have, therefore, the potential to reveal the connection between CRs and the postulated causes. Cosmic rays at these energies are almost entirely of Galactic origin and are expected to be isotropic due to the interaction with the Galactic magnetic field (GMF). Nevertheless, several theories have been postulated to explain the observations and the anisotropy can be imagined as a result of different astrophysical phenomena. An anisotropy can be induced by the heliospheric magnetic field. At lower energies around a TeV, the heliosphere may be able to induce a CR excess in the direction of the heliotail region (the so-called "tail-in" excess) \cite{Nagashima98}. At higher energies, the distribution of nearby recent supernova explosions has been postulated to be capable of creating a  large-scale anisotropy \cite{Erlykin06}. In addition to these effects Compton and Getting in 1935 \cite{CG35}  predicted a dipole effect due to the motion of the Earth with respect to an isotropic CR plasma rest frame. 
Two anisotropies can be postulated, one due to the Earth's motion around the Sun (Solar Dipole effect), that would appear in solar time,  and the other coming from the motion of the solar system around the galactic center (Galactic Compton Getting effect). This second effect would appear in sidereal time (ST) and it would be maximum with the CRs being at rest with respect to the
galactic center.\\
In this contribution we present the observations of the CR anisotropy by IceCube at two different angular scales. With this measurement we produced the first skymap in cosmic rays of the southern sky.

\section{DATASET}
The IceCube Neutrino Observatory currently under construction at the South Pole will be completed at the beginning of 2011 with 86 strings and 5,160 DOMs. While IceCube is designed to observe the sources of extragalactic neutrinos, it is also sensitive to TeV downward-going muons produced by cosmic ray air showers.  
Since the start of data taking in 2007, IceCube has recorded tens of billions of muon
events, and has accumulated the largest sample of TeV cosmic rays ever recorded in the southern hemisphere. The results presented here have been obtained analyzing the data collected in two periods, from June 2007 to March 2008 and from May 2008 to April 2009, in which IceCube was operating with 22 and 40 strings, respectively. The events used in the analysis have been reconstructed by an online likelihood-based reconstruction algorithm. The event rate is $\sim 240$ Hz for IceCube-22 and $\sim 750$ Hz for IceCube-40, with a median cosmic ray energy of about 20 TeV and a median angular resolution of $3^\circ$, where this value represent IceCube angular resolution for atmospheric muons with no quality cuts applied and has not to be confused with the angular resolution for neutrinos (better than 1$^\circ$ for IceCube-40).\\
These large statistics data sets allow for sensitive studies of the arrival direction distribution of the cosmic rays.

\section{DATA ANALYSIS AND RESULTS}

\subsection{Large scale anisotropy}

IceCube recently reported the first observation of the large angular scale anisotropy in arrival direction of CRs for the southern hemisphere \cite{Abbasi10}.\\
The analysis determines the relative intensity of the cosmic ray arrival direction in each declination band and it has to account for the spurious effects derived from the detector exposure asymmetries and non-uniform time coverage as well as for diurnal and seasonal variations of atmospheric conditions.
Fortunately, due to its favoured position at the South Pole, IceCube measurements of the sidereal variation is not affected by diurnal modulations because the whole sky is fully visible to the detector at any given time and because there is only one day and one night per year; on the other hand, the seasonal variation in the cosmic ray event rate is slow and does not affect the daily muon intensity significantly. \\
The remaining spurious effects which must be accounted for are an asymmetry in the detector geometry and non-uniformity in the time coverage of the data. The combinations of these two effects might induce an azimuthal asymmetry and mimic a sidereal anisotropy.
This effect is corrected normalizing the azimuthal distribution by re-weighting each event with the ratio between the average number of events and the number of events in the corresponding local azimuthal bin. Moreover, since the local azimuth distribution varies with the zenith angle, the sky is divided into four zenith bands with approximately the same number of events per band. The weighting is applied within each band to remove the detector asymmetry.

\begin{figure}[htb]
\includegraphics[height=50mm, width=75mm ]{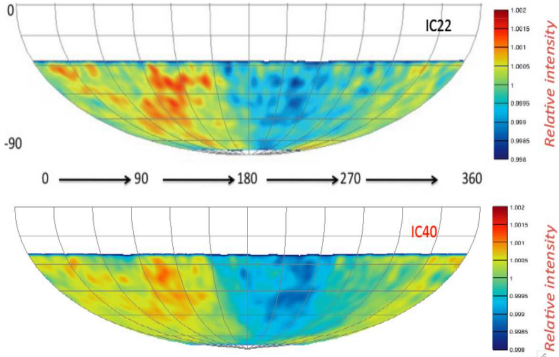}
\caption{Sky-map of the relative intensity in arrival direction of cosmic rays for IceCube-22 (top) \cite{Abbasi10}, and preliminary sky-map of the relative intensity for IceCube-40 (bottom), in equatorial coordinates. A gaussian smoothing
has been applied to the map for visualization purposes. Note that since the declination belts in the equatorial map are treated independently, the maps provide only information on the relative modulation of the arrival direction of cosmic rays along the right ascension.}
\label{fig:LargeScaleMaps}
\end{figure}

Figure \ref{fig:LargeScaleMaps} shows the relative intensity in arrival direction of the cosmic rays: on the top the relative intensity map obtained from the $\sim 4.6 \times 10^{9}$ events collected by IceCube-22 \cite{Abbasi10} and on the bottom the preliminary map obtained from the $\sim 12 \times 10^{9}$ events collected by IceCube-40.  Since the arrival direction distribution is dominated by the zenith angle dependence of the flux, the maps have been obtained by normalizing to unity each declination belt of $3^\circ$, which corresponds to the angular resolution of the data.
The two maps show the same anisotropy features and they both appear to be a continuation of the observed modulation in the northern hemisphere. 
To quantify the scale of the anisotropy the right ascension distribution has been fitted with a first and second-order harmonic function in the form of $\Sigma_{i=1}^{n=2} (A_i \times cos(i ((\alpha) - \phi_i))) +B$, where $A_i$ is the amplitude, $\phi_i$ is the phase and $B$ is a constant. The fit parameters as well as the quality of the fit for IceCube-22 are reported in \cite{Abbasi10}. 
\begin{figure}[htb]
\vspace{9pt}
\includegraphics[scale=0.46]{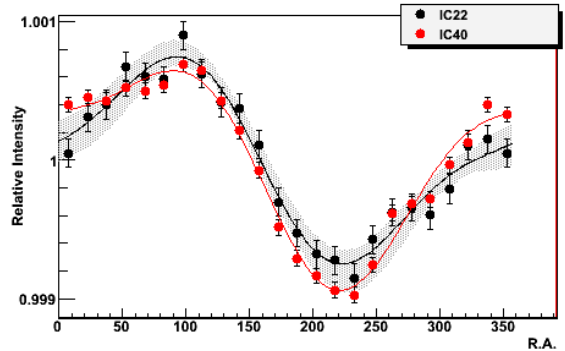}
\caption{Right ascension modulation (for $-72^\circ < \delta < -38^{\circ}$) of the relative intensity in arrival direction of cosmic rays for IceCube-22 (black dots) \cite{Abbasi10} and preliminary result for IceCube-40 (red dots). The data are shown with statistical uncertainties, and the lines represent the fit to the data. The gray band indicates the systematic uncertainties for IceCube-22.}
\label{fig:RelativeIntensity}
\end{figure}

Figure \ref{fig:RelativeIntensity} shows the right ascension modulation of the relative intensity in arrival direction of cosmic rays for IceCube-22 (black dots) \cite{Abbasi10} as well as the preliminary result for IceCube-40. The error bars represent the statistical uncertainties and the lines are the fit to the data. The gray band indicates the systematic uncertainties for IceCube-22 estimated taking into account daily and seasonal variations and non-uniform time coverage in the data.

\subsection{The Solar Dipole effect}

Any observer moving through a plasma of isotropic cosmic rays should observe a distinct difference in intensity between the direction of the velocity vector and the opposite direction. In 1935 Compton and Getting  predicted that the intensity of CRs should be observed like a dipole anisotropy \cite{CG35}; in this model an excess flux should appear with a maximum in right ascension between 290$^\circ$ and 340$^\circ$ and a minimum in right ascension between 110$^\circ$ and 160$^\circ$ \cite{Amenomori06}. 
The large scale anisotropy observed in IceCube (Fig.\ref{fig:RelativeIntensity}) can not be described as a pure dipole and the observed excess is not in the direction of motion of the solar system around the galaxy. We can conclude that the galactic Compton-Getting effect is at most one of several contributions to the sidereal anisotropy. \\
We also expect a dipole anisotropy caused by the Earth's motion around the Sun. The expected anisotropy is of order 10$^{-4}$ .
\begin{figure}[htb]
\vspace{9pt}
\includegraphics[height=55mm, width=80mm ]{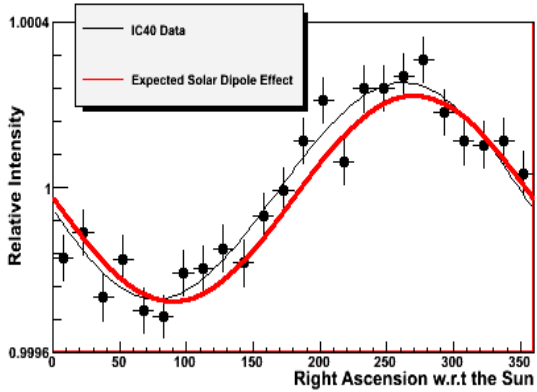}
\caption{Preliminary modulation of the cosmic ray relative intensity as a function of the relative arrival angle from the Sun in right ascension observed in IceCube-40. The black dots are the experimental data with the corresponding statistical error bars and the line represent the fit to the data. The red line represents the expected modulation due to Earth's motion.}
\label{fig:SolarDipole}
\end{figure}
If we represent the relative intensity as a function of the angular distance from the Sun in right ascension (defined as the difference between the right ascension of each event and the Sun position), we expect, over a full year, an excess in the direction of motion of the Earth around the Sun (at 270$^{\circ}$) and a minimum in the opposite direction. 
Figure \ref{fig:SolarDipole} shows the modulation observed in IceCube-40. The black dots represent the experimental data with the corresponding statistical uncertainties. The data have been fitted with a first-order harmonic function and the result of the fit is represented by the black line. The red line shows the expected solar dipole effect. In the plot the Sun is located at 0$^\circ$and the velocity vector of the Earth is at 270$^\circ$ which is where the maximum of the modulation is observed, as expected.\\
The dipole effect caused by Earth's motion around the Sun is visible only when the arrival directions are plotted in a frame where the Sun position is fixed in the sky. A fixed signal in this coordinate system is completely washed out in sidereal time over the course of one year. Therefore, the solar dipole effect is not visible in the sidereal skymap.

\subsection{Medium scale anisotropy}
Recent results from gamma ray experiments have shown evidence of intermediate ($\sim 10^\circ - 30^\circ$) scale structure in the CR arrival direction distribution.
Milagro observed two localized regions with flux excess of significance $> 10\sigma$ in the total data set of $2.2 \times 10^{11}$ events recorded over 7 years. The ``hot'' regions have fractional excesses of the order of several times $10^{-4}$ relative to the background \cite{Abdo08}. The same structures have been observed by ARGO-YBJ \cite{ARGOICRC09} and Tibet III \cite{TibetIII}. The origin of these excesses is still unknown but there seems to be some indications that the medium scale features might be part of the larger scale structure. \\
With the same technique used in the gamma ray field, IceCube can search for similar excess in the southern sky. The analysis technique is based on the {\em background estimation} from real data. The method can have different implementations and its general features are described in \cite{Alexandreas}. 
The analysis used in IceCube can be summarized in few steps: 
\begin{itemize}
\item
a signal map is created from the arrival directions of the events;
\item
a background map is calculated from the real data by randomly assigning detected event times to local arrival directions ({\em time scrambling}). The uncertainty in the background estimation is reduced by re-sampling the event time 20 times.  
\item 
the statistical significance of the result is calculated using the method of Li and Ma \cite{LiMa83}.
\end{itemize}

The length of the time window used to scramble the time of the events determines the maximum angular scale we can probe. Since we are interested in searching for features of angular scale $\sim 10^\circ - 30^\circ$ we choose a time window of 4 hours. In this way the time window is used to filter the signal coming from the large scale structure since  only the anisotropy features with angular scale smaller than $60^\circ$ are visible. 
Moreover, in order to maximize the sensitivity,  signal and background maps are smoothed on scales that correspond to the feature size.  \\
\begin{figure}[htb]
\vspace{9pt}
\includegraphics[height=45mm, width=80mm ]{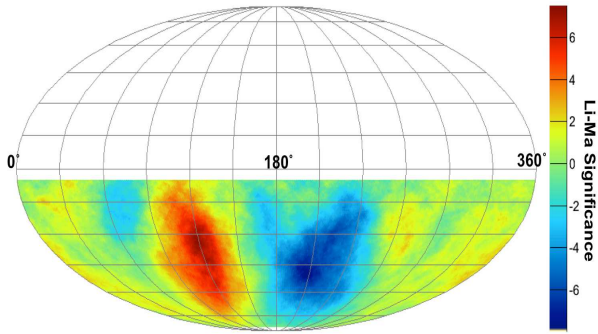}
\caption{Preliminary significance skymap in equatorial coordinates for a time window of 4 hours obtained with IceCube-40 data. A $25^\circ$ smoothing has been applied to improve the sensitivity to large features. The skymap shows a broad excess region around right ascension 120$^\circ$, with an equally strong deficit around right ascension 220$^\circ$.}
\label{fig:MediumScaleMap}
\end{figure}
Figure \ref{fig:MediumScaleMap} shows the preliminary significance skymap in equatorial coordinates for a time window of 4 hours obtained with $\sim 1.9 \times 10^{10}$ events collected in one year of IceCube-40. The map has been smoothed using an optimal radius of 25$^\circ$ (determined by examining the data) to improve the sensitivity to large features. The smoothing is performed adding counts from all pixels within a certain radius; due to this procedure the pixels in the map are strongly correlated. The skymap shows a broad excess region around right ascension 120$^\circ$, with an equally strong deficit around right ascension 220$^\circ$.\\
\begin{figure}[htb]
\vspace{9pt}
\includegraphics[height=55mm, width=85mm ]{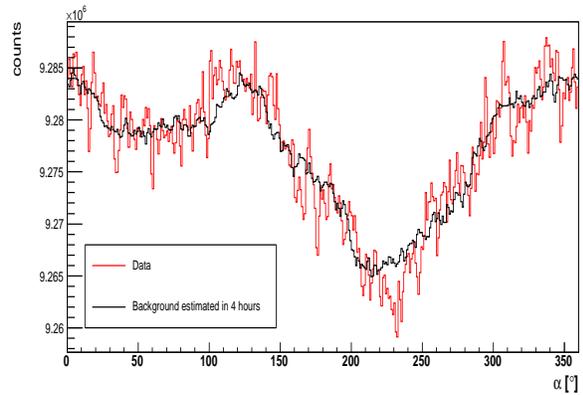}
\caption{Preliminary right ascension projection of signal and background events for $-45^\circ < \delta < -30^{\circ}$. The plot is made using independent $15^\circ \delta \times 1^\circ$ RA bins  (i.e. no smoothing applied). The data used here include only the full days to ensure a uniform exposure in RA. The two significant regions are already visible in the raw data.}
\label{fig:MediumScaleRAproj}
\end{figure}
Figure \ref{fig:MediumScaleRAproj} shows the preliminary RA projection of the number of events for the signal and the expected background for  $-45^\circ < \delta < -30^{\circ}$. The plot is made using independent $15^\circ \delta \times 1^\circ$ RA bins  (i.e. no smoothing applied). The data used for this figure has been chosen to include only full days in order to achieve an approximately uniform exposure as a function of RA. As can be seen, the background estimate as calculated via the time scrambling technique agrees well with the data. The excess and the deficit region shown in the skymap are visible in the raw data. 

\section{CONCLUSIONS}

IceCube observes, for the first time in the southern hemisphere, an anisotropy in the arrival direction of the galactic cosmic rays. The large scale structure seems to be a continuation of a previously observed modulation in the northern hemisphere. The IceCube skymap also reveals the presence of an anisotropy in intermediate scale ($\sim 10^\circ - 30^\circ$) with a broad excess region around right ascension 120$^\circ$, and an equally strong deficit around right ascension 220$^\circ$. The origin of these anisotropies is unknown. However there might be multiple superimposed causes, depending on the cosmic ray energy and the angular scale of the anisotropy. The postulated Compton-Getting effect \cite{CG35} should give rise to an energy independent dipole anisotropy with the maximum in the direction of the solar system and the deficit in the opposite direction. The measured modulation can not be described as a pure dipole and the observed excess is not in the direction of motion of the solar system around the galaxy; therefore, we can conclude that the galactic Compton-Getting effect, if present at all, is overshadowed by other effects. One could also imagine a scenario in which the cosmic ray excess might be associated with diffuse particle flows coming from nearby cosmic ray sources like Vela SNR. 
Recently the MILAGRO excess has been interpreted by invoking Geminga pulsar as a possible source  \cite{Salvati08}, \cite{Drury08}. However, it is likely that a localized excess of multi-TeV CRs is originated at close distances. In Lazarian \& Desiati it is proposed that both the localized excess observed by Milagro toward the direction of the heliotail (i.e. the portion of the heliosphere downstream the interstellar wind) and the broad excess of cosmic rays observed at sub-TeV energies from the same direction (tail-in excess) \cite{Nagashima98}  are generated by first-order Fermi acceleration via stochastic magnetic reconnection in the heliotail up to about 10 TeV \cite{Lazarian10}. Reconnection is generated when solar magnetic polarity reversal regions due to the 11-year solar cycles are compressed by the solar wind in the heliotail. In the reconnection regions, magnetic mirrors form where acceleration might be efficient to distort spectrum and arrival direction of the cosmic rays.

\end{document}